%% file: main.tex
\pdfoutput=1

\documentclass[12pt,a4paper]{article}

\usepackage{ifthen} 
\newboolean{pdflatex}
\setboolean{pdflatex}{true} 

\newboolean{articletitles}
\setboolean{articletitles}{true} 

\newboolean{uprightparticles}
\setboolean{uprightparticles}{false} 

\def\paperauthors{LHCb collaboration} 
\def\paperasciititle{Observation of an excited B_c+ state} 
\def\papertitle{Observation of an excited $B_c^+$ state} 
\def\paperkeywords{{High Energy Physics}, {LHCb}} 
\def\papercopyright{\the\year\ CERN for the benefit of the LHCb collaboration} 
\def\paperlicence{CC-BY-4.0 licence}
\def\paperlicenceurl{https://creativecommons.org/licenses/by/4.0/}

\input{preamble}
\input{mydefine}

\usepackage{longtable} 
\usepackage{subfig}

\begin{document}

\renewcommand{\thefootnote}{\fnsymbol{footnote}}
\setcounter{footnote}{1}

\input{title-LHCb-PAPER}


\renewcommand{\thefootnote}{\arabic{footnote}}
\setcounter{footnote}{0}



\pagestyle{plain} 
\setcounter{page}{1}
\pagenumbering{arabic}


%

\input{Bc2S-body}

\input{acknowledgements}

\addcontentsline{toc}{section}{References}
\bibliographystyle{LHCb}
\bibliography{main} 
 
\newpage
\input{LHCb_Authorship_flat_07-Feb-2019.tex}

\end{document}

%% file: preamble.tex
\usepackage[top=1in, bottom=1.25in, left=1in, right=1in]{geometry}

\usepackage{microtype}
\usepackage{lineno}  
\usepackage{xspace} 
\usepackage{caption} 

\usepackage{graphicx}  
\usepackage{color}
\usepackage{colortbl}
\graphicspath{{./figs/}} 
\DeclareGraphicsExtensions{.pdf,.PDF,png,.PNG}

\usepackage{amsmath} 
\usepackage{amssymb}
\usepackage{amsfonts}
\usepackage{upgreek} 

\newcommand*\patchAmsMathEnvironmentForLineno[1]{%
\expandafter\let\csname old#1\expandafter\endcsname\csname #1\endcsname
\expandafter\let\csname oldend#1\expandafter\endcsname\csname
end#1\endcsname
 \renewenvironment{#1}%
   {\linenomath\csname old#1\endcsname}%
   {\csname oldend#1\endcsname\endlinenomath}%
}
\newcommand*\patchBothAmsMathEnvironmentsForLineno[1]{%
  \patchAmsMathEnvironmentForLineno{#1}%
  \patchAmsMathEnvironmentForLineno{#1*}%
}
\AtBeginDocument{%
\patchBothAmsMathEnvironmentsForLineno{equation}%
\patchBothAmsMathEnvironmentsForLineno{align}%
\patchBothAmsMathEnvironmentsForLineno{flalign}%
\patchBothAmsMathEnvironmentsForLineno{alignat}%
\patchBothAmsMathEnvironmentsForLineno{gather}%
\patchBothAmsMathEnvironmentsForLineno{multline}%
\patchBothAmsMathEnvironmentsForLineno{eqnarray}%
}

\usepackage{hyperxmp}

\usepackage[pdftex,
            pdfauthor={\paperauthors},
            pdftitle={\paperasciititle},
            pdfkeywords={\paperkeywords},
            pdfcopyright={Copyright (C) \papercopyright},
            pdflicenseurl={\paperlicenceurl}]{hyperref}

\usepackage[colorinlistoftodos,textsize=scriptsize]{todonotes}

\usepackage[all]{hypcap} 

\input{lhcb-symbols-def} 

\usepackage{cite} 
\usepackage{mciteplus}

%% file: lhcb-symbols-def.tex
\usepackage{xspace} 
\usepackage{upgreek}

\def\lhcb   {\mbox{LHCb}\xspace}
\def\atlas  {\mbox{ATLAS}\xspace}

\def\cdf    {\mbox{CDF}\xspace}

\def\tevatron {Tevatron\xspace}

\def\MagUp {\mbox{\em Mag\kern -0.05em Up}\xspace}

\ifthenelse{\boolean{uprightparticles}}%
{

 \def\Ppi         {\ensuremath{\uppi}\xspace}

 \def\Ppsi        {\ensuremath{\uppsi}\xspace}

 \def\PDelta      {\ensuremath{\Delta}\xspace}                 
 \def\PXi         {\ensuremath{\Xi}\xspace}                 
 \def\PLambda     {\ensuremath{\Lambda}\xspace}                 
 \def\PSigma      {\ensuremath{\Sigma}\xspace}                 
 \def\POmega      {\ensuremath{\Omega}\xspace}                 
 \def\PUpsilon    {\ensuremath{\Upsilon}\xspace}

 \def\PB      {\ensuremath{\mathrm{B}}\xspace}                 
                  
 \def\PD      {\ensuremath{\mathrm{D}}\xspace}

 \def\PJ      {\ensuremath{\mathrm{J}}\xspace}                 
 \def\PK      {\ensuremath{\mathrm{K}}\xspace}

 \def\Pb      {\ensuremath{\mathrm{b}}\xspace}                 
 \def\Pc      {\ensuremath{\mathrm{c}}\xspace}

 \def\Pi      {\ensuremath{\mathrm{i}}\xspace}

 \def\thebaroffset{0.0em}
}
{

 \def\Ppi         {\ensuremath{\pi}\xspace}

 \def\Ppsi        {\ensuremath{\psi}\xspace}                 
                  
 \mathchardef\PDelta="7101
 \mathchardef\PXi="7104
 \mathchardef\PLambda="7103
 \mathchardef\PSigma="7106
 \mathchardef\POmega="710A
 \mathchardef\PUpsilon="7107
                  
 \def\PB      {\ensuremath{B}\xspace}                 
                  
 \def\PD      {\ensuremath{D}\xspace}

 \def\PJ      {\ensuremath{J}\xspace}                 
 \def\PK      {\ensuremath{K}\xspace}

 \def\Pb      {\ensuremath{b}\xspace}                 
 \def\Pc      {\ensuremath{c}\xspace}

 \def\Pi      {\ensuremath{i}\xspace}

 \def\thebaroffset{0.18em}
}
\newcommand{\offsetoverline}[2][\thebaroffset]{\kern #1\overline{\kern -#1 #2}}%

\makeatletter
\ifcase \@ptsize \relax
  \newcommand{\miniscule}{\@setfontsize\miniscule{4}{5}}
\or
  \newcommand{\miniscule}{\@setfontsize\miniscule{5}{6}}
\or
  \newcommand{\miniscule}{\@setfontsize\miniscule{5}{6}}
\fi
\makeatother

\DeclareRobustCommand{\optbar}[1]{\shortstack{{\miniscule (\rule[.5ex]{1.25em}{.18mm})}
  \\ [-.7ex] $#1$}}

\def\cquark    {{\ensuremath{\Pc}}\xspace}
\def\cquarkbar {{\ensuremath{\overline \cquark}}\xspace}

\def\bquark    {{\ensuremath{\Pb}}\xspace}
\def\bquarkbar {{\ensuremath{\overline \bquark}}\xspace}

\def\pion   {{\ensuremath{\Ppi}}\xspace}

\def\pip    {{\ensuremath{\pion^+}}\xspace}
\def\pim    {{\ensuremath{\pion^-}}\xspace}

\def\kaon    {{\ensuremath{\PK}}\xspace}

\def\KorKbar {\kern \thebaroffset\optbar{\kern -\thebaroffset \PK}{}\xspace}

\def\KS      {{\ensuremath{\kaon^0_{\mathrm{S}}}}\xspace}

\def\DorDbar {\kern \thebaroffset\optbar{\kern -\thebaroffset \PD}\xspace}

\def\B       {{\ensuremath{\PB}}\xspace}

\def\BorBbar {\kern \thebaroffset\optbar{\kern -\thebaroffset \PB}\xspace}

\def\Bc      {{\ensuremath{\B_\cquark^+}}\xspace}
\def\Bcp     {{\ensuremath{\B_\cquark^+}}\xspace}

\def\jpsi     {{\ensuremath{{\PJ\mskip -3mu/\mskip -2mu\Ppsi\mskip 2mu}}}\xspace}

\def\Y#1S{\ensuremath{\PUpsilon{(#1S)}}\xspace}

\def\LorLbar     {\kern \thebaroffset\optbar{\kern -\thebaroffset \PLambda}\xspace}

\newcommand{\decay}[2]{\mbox{\ensuremath{#1\!\to #2}}\xspace}         

\def\to                 {\ensuremath{\rightarrow}\xspace}

\def\AT#1     {\ensuremath{A_{\mathrm{T}}^{#1}}\xspace}           

\def\C#1      {\ensuremath{\mathcal{C}_{#1}}\xspace}                       
\def\Cp#1     {\ensuremath{\mathcal{C}_{#1}^{'}}\xspace}                    
\def\Ceff#1   {\ensuremath{\mathcal{C}_{#1}^{\mathrm{(eff)}}}\xspace}        
\def\Cpeff#1  {\ensuremath{\mathcal{C}_{#1}^{'\mathrm{(eff)}}}\xspace}       
\def\Ope#1    {\ensuremath{\mathcal{O}_{#1}}\xspace}                       
\def\Opep#1   {\ensuremath{\mathcal{O}_{#1}^{'}}\xspace}                    

\newcommand{\nospaceunit}[1]{\ensuremath{\text{#1}}}       
\newcommand{\aunit}[1]{\ensuremath{\text{\,#1}}}       

\newcommand{\tev}{\aunit{Te\kern -0.1em V}\xspace}
\newcommand{\gev}{\aunit{Ge\kern -0.1em V}\xspace}
\newcommand{\mev}{\aunit{Me\kern -0.1em V}\xspace}
\newcommand{\kev}{\aunit{ke\kern -0.1em V}\xspace}
\newcommand{\ev}{\aunit{e\kern -0.1em V}\xspace}
\newcommand{\mevc}{\ensuremath{\aunit{Me\kern -0.1em V\!/}c}\xspace}
\newcommand{\gevc}{\ensuremath{\aunit{Ge\kern -0.1em V\!/}c}\xspace}
\newcommand{\mevcc}{\ensuremath{\aunit{Me\kern -0.1em V\!/}c^2}\xspace}
\newcommand{\gevcc}{\ensuremath{\aunit{Ge\kern -0.1em V\!/}c^2}\xspace}

\def\mum  {\ensuremath{\,\upmu\nospaceunit{m}}\xspace}

\def\fb   {\ensuremath{\aunit{fb}}\xspace}
\def\invfb   {\ensuremath{\fb^{-1}}\xspace}

\def\ps   {\ensuremath{\aunit{ps}}\xspace}

\newcommand{\stat}{\aunit{(stat)}\xspace}
\newcommand{\syst}{\aunit{(syst)}\xspace}

\newcommand{\chisq}{\ensuremath{\chi^2}\xspace}

\newcommand{\chisqip}{\ensuremath{\chi^2_{\text{IP}}}\xspace}

\def\gsim{{~\raise.15em\hbox{$>$}\kern-.85em
          \lower.35em\hbox{$\sim$}~}\xspace}
\def\lsim{{~\raise.15em\hbox{$<$}\kern-.85em
          \lower.35em\hbox{$\sim$}~}\xspace}

\def\sqs   {\ensuremath{\protect\sqrt{s}}\xspace}

\def\pt         {\ensuremath{p_{\mathrm{T}}}\xspace}

\def\bcvegpy    {\mbox{\textsc{BcVegPy}}\xspace}

\def\evtgen     {\mbox{\textsc{EvtGen}}\xspace}

\def\geant      {\mbox{\textsc{Geant4}}\xspace}

\def\photos     {\mbox{\textsc{Photos}}\xspace}

\def\pythia     {\mbox{\textsc{Pythia}}\xspace}

\xspace

\def\tell1  {TELL1\xspace}
\def\ukl1   {UKL1\xspace}

\newcommand{\ie}{\mbox{\itshape i.e.}\xspace}

%% file: mydefine.tex
\newcommand{\xx}{\ensuremath{\kern 0.5em }}

\newcommand{\statmone}{\ensuremath{6841.1\pm0.6\stat\pm0.8\,(\Bcp) \mevcc}}
\newcommand{\statmtwo}{\ensuremath{6872.1\pm1.3\stat\pm0.8\,(\Bcp) \mevcc}}
\newcommand{\statmgap}{\ensuremath{31.1 \pm 1.4\stat \mevcc}}

\newcommand{\Qone}{\ensuremath{566.3 \pm 0.6\stat \pm 0.1\syst \mevcc}}
\newcommand{\Qtwo}{\ensuremath{597.2 \pm 1.3\stat \pm 0.1\syst \mevcc}}
\newcommand{\mone}{\ensuremath{6841.2 \pm 0.6\stat \pm 0.1\syst \pm 0.8\,(\Bcp)\mevcc}}
\newcommand{\mtwo}{\ensuremath{6872.1 \pm 1.3\stat \pm 0.1\syst \pm 0.8\,(\Bcp)\mevcc}}
\newcommand{\mgap}{\ensuremath{31.0 \pm 1.4\stat \pm 0.0\syst \mevcc}}

\def\Bcstar    {{\ensuremath{\B_\cquark^{*+}}}\xspace}
\def\Bctriplet  {{\ensuremath{\B_\cquark(1^{3}S_{1})^+}}\xspace}
\def\Bctwos      {{\ensuremath{\B_\cquark(2S)^+}}\xspace}
\def\Bctwosstar  {{\ensuremath{\B_\cquark^{*}(2S)^+}}\xspace}
\def\Bctwossinglet      {{\ensuremath{\B_\cquark(2^{1}S_{0})^+}}\xspace}
\def\Bctwostriplet  {{\ensuremath{\B_\cquark(2^{3}S_{1})^+}}\xspace}

\def\twoBctwos {{\ensuremath{\B_\cquark^{(*)}}(2S)^+}\xspace}

\def\MBctwos    {\ensuremath{M(\Bctwos)}}
\def\MBctwosstar    {\ensuremath{M(\Bctwosstar)}}

\def\Mrec {\ensuremath{M(\Bctwosstar)_{\rm rec}}}

\usepackage{multirow} 
\usepackage{booktabs} 
\usepackage{rotating}

%% file: title-LHCb-PAPER.tex

\begin{titlepage}
\pagenumbering{roman}

\vspace*{-1.5cm}
\centerline{\large EUROPEAN ORGANIZATION FOR NUCLEAR RESEARCH (CERN)}
\vspace*{1.5cm}
\noindent
\begin{tabular*}{\linewidth}{lc@{\extracolsep{\fill}}r@{\extracolsep{0pt}}}
\ifthenelse{\boolean{pdflatex}}
{\vspace*{-1.5cm}\mbox{\!\!\!\includegraphics[width=.14\textwidth]{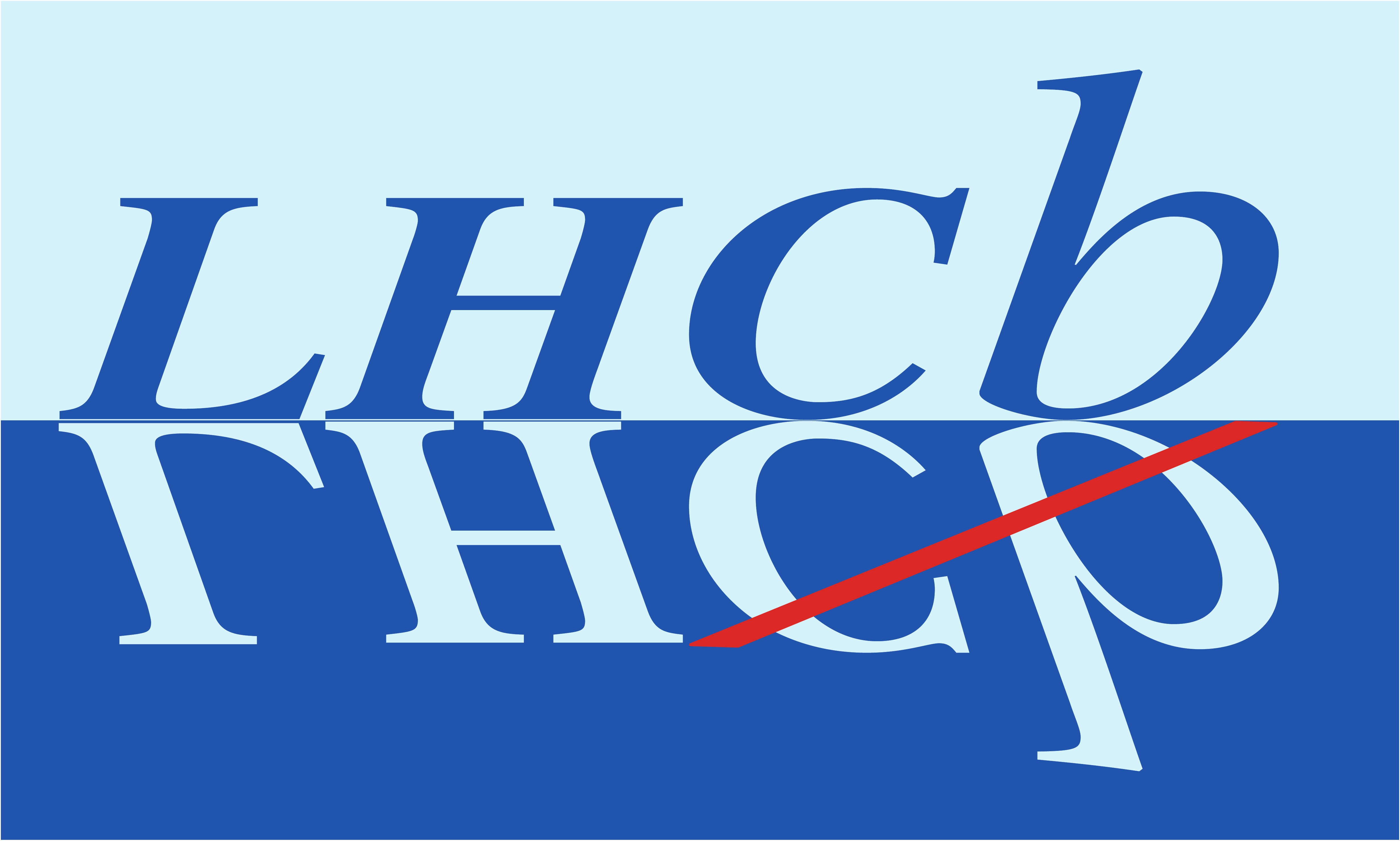}} & &}%
{\vspace*{-1.2cm}\mbox{\!\!\!\includegraphics[width=.12\textwidth]{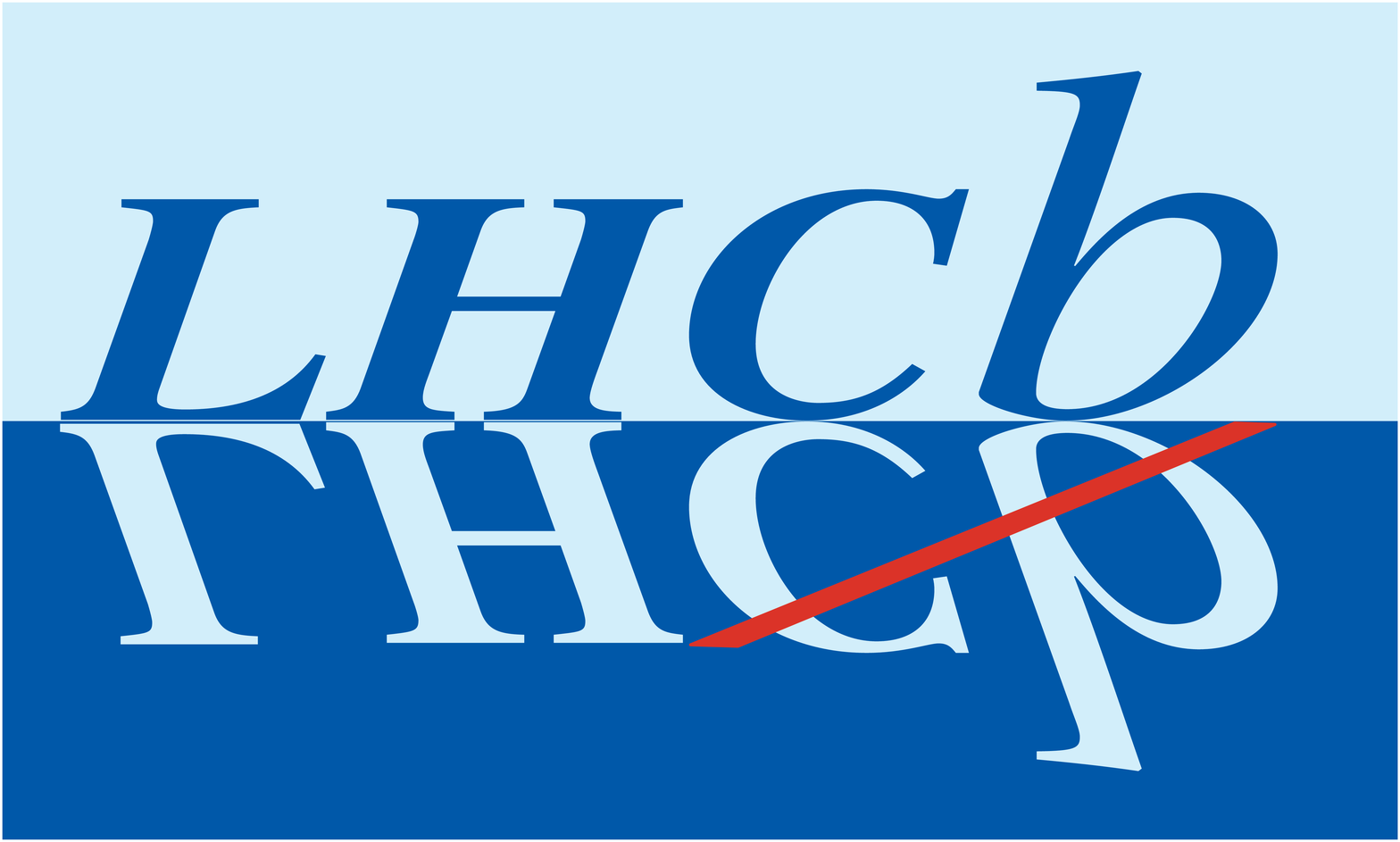}} & &}%
\\
 & & CERN-EP-2019-050 \\  
 & & LHCb-PAPER-2019-007 \\  
 & & June 12, 2019 \\ 
 & & \\
\end{tabular*}

\vspace*{4.0cm}

{\normalfont\bfseries\boldmath\huge
\begin{center}
  \papertitle 
\end{center}
}

\vspace*{2.0cm}

\begin{center}
\paperauthors\footnote{Authors are listed at the end of this Letter.}
\end{center}

\vspace{\fill}

\begin{abstract}
  \noindent
  Using $pp$ collision data corresponding to an integrated luminosity of
  $8.5\,\mathrm{fb}^{-1}$ recorded by the LHCb experiment 
  at centre-of-mass energies of $\sqrt{s} = 7$, $8$ and $13\mathrm{\,Te\kern -0.1em V}$,
  the observation of an excited $B_c^+$ state in the $B_c^+\pi^+\pi^-$ invariant-mass spectrum is reported.
  The observed peak has a mass of
  $6841.2 \pm 0.6 {\,\rm (stat)\,} \pm 0.1 {\,\rm (syst)\,} \pm 0.8\,(B_c^+) \mathrm{\,MeV}/c^2$,
  where the last uncertainty is due to the limited knowledge of the $B_c^+$ mass.
  It is consistent with expectations of the $B_c^{*}(2^{3}S_{1})^+$ state reconstructed
  without the low-energy photon from the $B_c^{*}(1^{3}S_{1})^+ \to B_c^+ \gamma$ decay
  following $B_c^{*}(2^{3}S_{1})^+ \to B_c^{*}(1^{3}S_{1})^+ \pi^+ \pi^-$.
  A second state is seen with a global (local) statistical significance
  of $2.2\,\sigma$ ($3.2\,\sigma$) and
  a mass of
  $6872.1 \pm 1.3 {\,\rm (stat)\,} \pm 0.1 {\,\rm (syst)\,} \pm 0.8\,(B_c^+) \mathrm{\,MeV}/c^2$,
  and is consistent with the $B_c(2^{1}S_{0})^+$ state.
  These mass measurements are the most precise to date.
\end{abstract}

\vspace*{2.0cm}

\begin{center}
  Published in Phys.~Rev.~Lett. 122 (2019) 232001
\end{center}

\vspace{\fill}

{\footnotesize 
\centerline{\copyright~\papercopyright. \href{\paperlicenceurl}{\paperlicence}.}}
\vspace*{2mm}

\end{titlepage}


\newpage
\setcounter{page}{2}
\mbox{~}
\newpage

\cleardoublepage

%% file: Bc2S-body.tex
The $B_c$ meson family is unique in the Standard Model
as its states are formed from two heavy quarks of different flavours.
The spectrum of masses of $B_c$ mesons can reveal information on heavy-quark dynamics
and improve the understanding of the strong interaction.
Specifically,
it provides tests of
 nonrelativistic quark-potential models~\cite{Gershtein:1987jj, Chen:1992fq, Eichten:1994gt, Kiselev:1994rc,
  Gupta:1995ps, Fulcher:1998ka, Ebert:2002pp, Godfrey:2004ya, Wei:2010zza, El:2005, Zeng:1994vj, Davies:1996gi, Dowdall:2012ab}, 
which have been successfully applied to quarkonium,
since the $B_c$ family shares properties with both the charmonium and bottomonium systems.
The $B_c$ family is predicted to have a rich spectroscopy
by various potential models~\cite{Gershtein:1987jj, Chen:1992fq, Eichten:1994gt, Kiselev:1994rc,
  Gupta:1995ps, Fulcher:1998ka, Ebert:2002pp, Godfrey:2004ya, Wei:2010zza, El:2005, Zeng:1994vj, Davies:1996gi, Dowdall:2012ab}
and lattice quantum chromodynamics~\cite{Davies:1996gi}.
However, the $B_c$ mesons are much less explored compared to quarkonia due to the small production rate,
since their predominant production mechanism 
requires the production of both $\cquark\cquarkbar$ and $\bquark\bquarkbar$ pairs.
The ground state meson, $\Bcp$,
was first observed by the \cdf experiment~\cite{Abe:1998wi,*Abe:1998fb} at the \tevatron collider.
Knowledge of the properties of the $\Bcp$ meson has been greatly advanced by the 
\lhcb experiment
with the measurement of the $\Bcp$ mass, lifetime and production rate~\cite{
LHCb-PAPER-2012-028,LHCb-PAPER-2013-010,LHCb-PAPER-2014-039,LHCb-PAPER-2013-063,LHCb-PAPER-2014-060,LHCb-PAPER-2014-050},
and the discovery and precise measurement of the branching fractions of several new decay channels~\cite{
LHCb-PAPER-2011-044,LHCb-PAPER-2012-054,LHCb-PAPER-2013-010,LHCb-PAPER-2013-021,LHCb-PAPER-2013-044,LHCb-PAPER-2013-047,LHCb-PAPER-2014-009, LHCb-PAPER-2015-024,LHCb-PAPER-2016-020,LHCb-PAPER-2016-055,LHCb-PAPER-2016-058}. Charge conjugation is implied throughout this Letter.

\begin{figure}[htb]
 \begin{center}
    \includegraphics*[width=.6\textwidth]{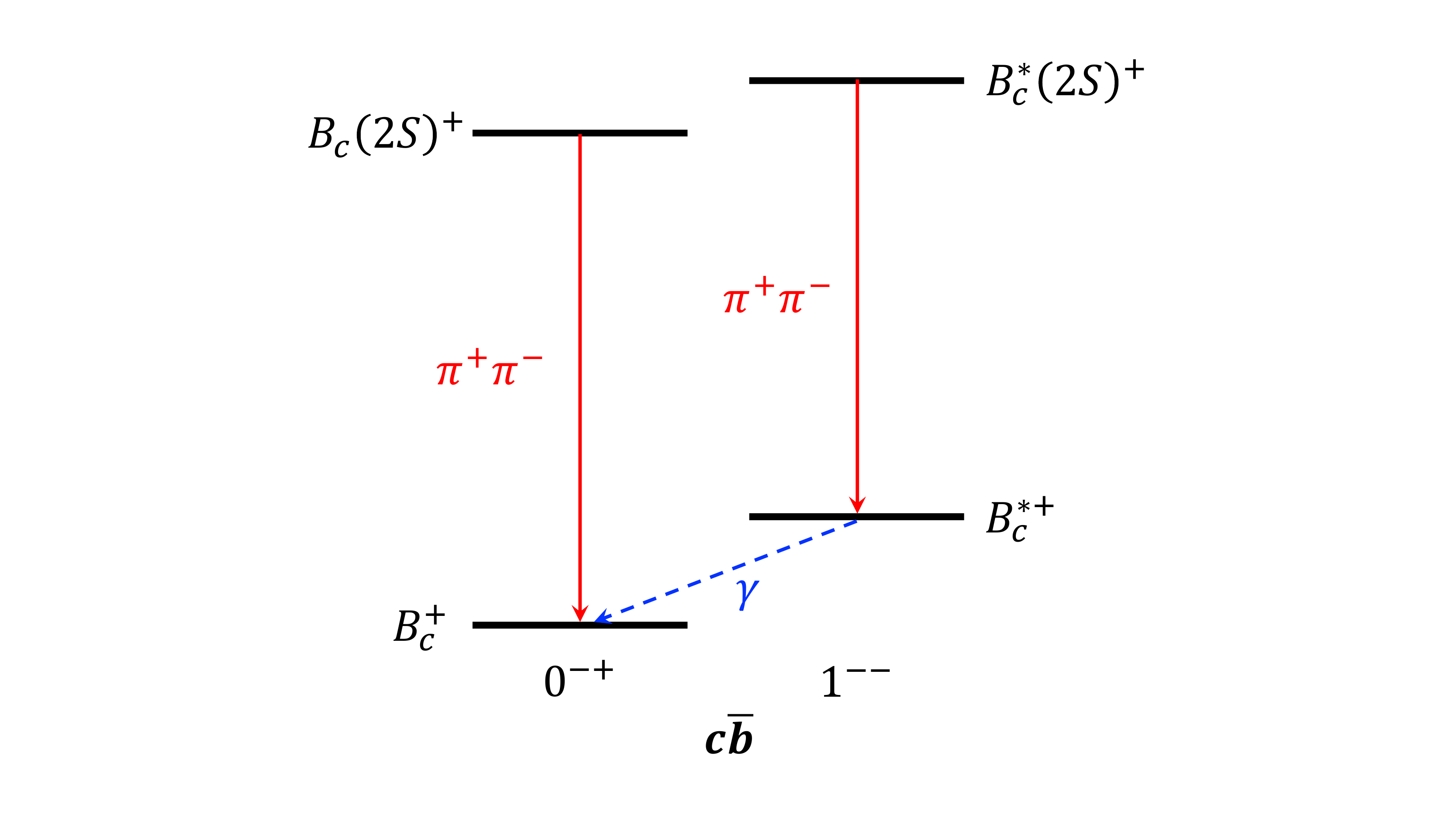}
   \caption{Transitions among the $\twoBctwos$ and $B_c^{(*)+}$ states. 
   }
   \label{fig:Bc2Sdecay}
 \end{center}
\end{figure}

Excited $\Bcp$ states that 
lie below the threshold for decay into a beauty and charm meson pair
are expected to have decay widths smaller than a few hundred $\kev$~\cite{Eichten:1994gt,Kiselev:1994rc}. 
Depending on its mass, an excited $\Bcp$ resonance may undergo either cascade radiative or pionic decays to the $\Bcp$ state,
which decays weakly.
The second $S$-wave $B_c$ state
occurs as either a 
pseudoscalar $(0^-)$
or a vector $(1^-)$ spin state, \ie, 
the singlet $\Bctwossinglet$ or the triplet $\Bctwostriplet$. 
The $\Bctwossinglet$ and $\Bctwostriplet$ states are denoted as
$\Bctwos$ and $\Bctwosstar$, respectively.
The $\Bctwos$ state decays directly to $\Bcp\pip\pim$,
while the $\Bctwosstar$ state decays to $\Bctriplet\pip\pim$, 
followed by the $\Bctriplet \to \Bcp \gamma$ electromagnetic transition.
The low-energy photon produced in this decay is not considered in this analysis,
since the reconstruction efficiency for such photons is too low
to be useful with the current data sample.
The $\Bctriplet$ state is denoted as $\Bcstar$ hereafter.
The transitions among the $\twoBctwos$ and $B_c^{(*)+}$ states are illustrated in Fig.~\ref{fig:Bc2Sdecay}.
Decays of both $\twoBctwos$ states produce a narrow peak in the $\Bcp\pip\pim$ invariant-mass spectrum~\cite{Gao:2010zzc,Berezhnoy:2013sla}, 
however, the $\Bctwosstar$ state peaks at ${\Mrec = \MBctwosstar - \Delta M(\Bcstar)}$ due to the missing photon,
where $\Delta M(\Bcstar)$ is the mass difference between 
the intermediate state $\Bcstar$ and the $\Bcp$ meson.
Since the $\Bcstar$ state has not been observed yet, 
the quantity $\Delta M(\Bcstar)$ is unknown 
and the value of $\MBctwosstar$ can not be determined with this technique at the moment.
Taking into account the unreconstructed photon, the mass difference between the two peaks in the $\Bcp\pip\pim$ mass distribution
originating from the two $\twoBctwos$ states, 
$\MBctwos - \Mrec$, 
is predicted to be in the range 11 to 53$\mevcc$~\cite
{Gershtein:1987jj, Chen:1992fq, Eichten:1994gt, Kiselev:1994rc,Gupta:1995ps, Fulcher:1998ka, Ebert:2002pp, Godfrey:2004ya, Wei:2010zza, El:2005, Zeng:1994vj, Davies:1996gi, Dowdall:2012ab}.
The production cross-section of the $\Bctwosstar$ state 
is predicted to be twice as large as
that of the $\Bctwos$ state~\cite{Chang:2015qea,Gouz:2002kk,Godfrey:2004ya,Gao:2010zzc},
while the branching fractions of the decays \mbox{$\decay{\Bctwos}{\Bcp\pip\pim}$} and
\mbox{$\decay{\Bctwosstar}{\Bcstar\pip\pim}$} are expected to be similar~\cite{Gouz:2002kk,Godfrey:2004ya}.

With the large samples of $\Bcp$ mesons produced at the Large Hadron Collider, the \atlas collaboration 
first reported the observation of a signal in the $\Bcp\pip\pim$ mass distribution
peaking at a value of \mbox{$6842 \pm 4\stat \pm 5\syst \mevcc$}
using $pp$ collision data
at $\sqs = 7$ and $8\tev$ corresponding to a luminosity of $24\invfb$~\cite{Aad:2014laa}. 
Due to large mass resolution and low signal yield, 
no determination could be made as to 
whether the observed peak was either the $\Bctwos$, the $\Bctwosstar$ state, or a combination of the two states.
The \lhcb experiment also performed a search for excited $\Bc$ states 
in the $\Bcp\pip\pim$ mass distribution
using $pp$ collision data
at centre-of-mass energy of $\sqs = 8\tev$,
corresponding to an integrated luminosity of $2\invfb$.
No evidence of any signal was found~\cite{LHCb-PAPER-2017-042}.
Recently, the CMS collaboration reported the observation of the
$\Bctwos$ and $\Bctwosstar$ states~\cite{Sirunyan:2019osb},
in which the mass of the $\Bctwos$ state 
and the mass difference between the two peaks
were measured to be
$6871.0\pm1.2\stat\pm0.8\syst\pm0.8(\Bcp)\mevcc$
and $29.0\pm1.5\stat\pm0.7\syst\mevcc$,
respectively.
The third uncertainty is due to the limited knowledge of the $\Bcp$ mass.

This Letter presents an updated search for excited $B_c$ mesons in the $\Bcp\pip\pim$ mass distribution.
The analysis makes use of Run 1 and Run 2 data collected by the \lhcb experiment
from 2011 to 2018 at centre-of-mass energies of $\sqs = 7$, $8$ and $13\tev$,
corresponding to integrated luminosities of about $1.0$, $2.0$ and
$5.5\invfb$, respectively.

The \lhcb detector~\cite{Alves:2008zz,LHCb-DP-2014-002} is a
single-arm forward spectrometer covering the pseudorapidity range $2 < \eta < 5$, designed for
the study of particles containing \bquark\ and/or \cquark\ quarks. 
The detector elements that are particularly
relevant to this analysis are: 
a silicon-strip vertex detector surrounding the $pp$ interaction
region that allows \cquark\ and \bquark\ hadrons to be identified from their characteristically long
flight distance; 
a tracking system that provides a measurement of the momentum, $p$, of charged particles with
a relative uncertainty that varies from 0.5\% at low momentum to 1.0\% at 200\gevc;
and two ring-imaging Cherenkov detectors that are able to discriminate between
different species of charged hadrons.
The minimum distance of a track to a primary vertex (PV), the impact parameter (IP),
is measured with a resolution of $(15+29/\pt)\mum$,
where \pt is the component of the momentum transverse to the beam, in\,\gevc.
The online event selection is performed by a trigger, 
which consists of a hardware stage, based on information from the calorimeter and muon
systems, followed by a software stage, which applies a full event
reconstruction.
At the hardware stage, 
events are required to have at least one muon with high transverse momentum, $\pt$, or 
a hadron with high transverse energy. 
At the software stage, 
two muon tracks or three charged tracks 
are required to have high \pt and to form a secondary vertex
with a significant displacement from the interaction point.
The momentum scale in data is calibrated using 
the $\jpsi$ and $B^+$ mesons~\cite{LHCb-PAPER-2011-035}
with well-known masses.

Simulated samples are used 
to model the signal behaviour. 
In the simulation, $pp$ collisions are generated using
\pythia6~\cite{Sjostrand:2006za} 
 with a specific \lhcb
configuration~\cite{LHCb-PROC-2010-056}.  
The generator \bcvegpy~\cite{Chang:2015qea}
is used to simulate the production of $\Bcp$ mesons.
Decays of unstable particles
are described by \evtgen~\cite{Lange:2001uf}, in which final-state
radiation is generated using \photos~\cite{Golonka:2005pn}. 
The interaction of the generated particles with the detector, and its response,
are implemented using the \geant
toolkit~\cite{Allison:2006ve} as described in
Ref.~\cite{LHCb-PROC-2011-006}.

To form the $\twoBctwos$ candidates,
first the intermediate $\Bcp$ state is reconstructed from
the $\Bcp\to\jpsi\pip$ decay.
The $\jpsi$ candidates are reconstructed with 
a pair of oppositely charged particles identified as muons.
The muons are required to have
$\pt>550\mevc$ and
good track-fit quality.
They are required to form a common decay vertex
with an invariant mass in the range $[3040,3140]\mevcc$,
corresponding to approximately six
times the $\jpsi$ mass resolution. 
The $\jpsi$ candidate is combined with a charged pion to form the $\Bcp$ candidate.
Each particle is associated to the PV that has the smallest value of $\chisqip$,
where $\chisqip$ is defined as the difference in the vertex-fit $\chisq$ of a given PV
reconstructed with and without the particle under consideration.
The pion must have $\pt>1000\mevc$,
good track-fit quality,
and be inconsistent with originating from any PV.
The $\Bcp$ candidate is required to 
have a good-quality vertex,
a trajectory consistent with coming from its associated PV,
and a decay time larger than $0.2\ps$.

To further suppress background,
a boosted decision tree~(BDT)~\cite{Breiman,AdaBoost} classifier 
is used, as done in the $\Bcp$ production measurement~\cite{LHCb-PAPER-2014-050}.
The input variables of the BDT classifier are taken to be
the $\pt$ of each muon, the $\jpsi$ meson
and the charged pion;
the decay length, decay time and vertex-fit $\chisq$ of the $\Bcp$ meson;
and the $\chisqip$ of the muons, the pion,
the $\jpsi$ meson and the $\Bcp$ meson
with respect to the associated PV.
The BDT classifier is trained using signal candidates from simulation
and background candidates from the upper sideband of the $\jpsi\pip$ mass distribution in data,
corresponding to the range $[6370,6600]\mevcc$.
The BDT threshold is chosen to maximise
$S/\sqrt{S+B}$,
where $S$ and $B$ are the expected yields of signal and background
in the range $M(\jpsi\pip)\in[6251,6301]\mevcc$, respectively.
This mass window corresponds to around four times the resolution of $M(\jpsi\pip)$.
To improve the signal-to-background ratio in the $\twoBctwos$ search,
the transverse momentum of the $\Bcp$ meson 
is required to be larger than $10\gevc$.

\begin{figure}[tb]
 \begin{center}
    \includegraphics*[width=.49\textwidth]{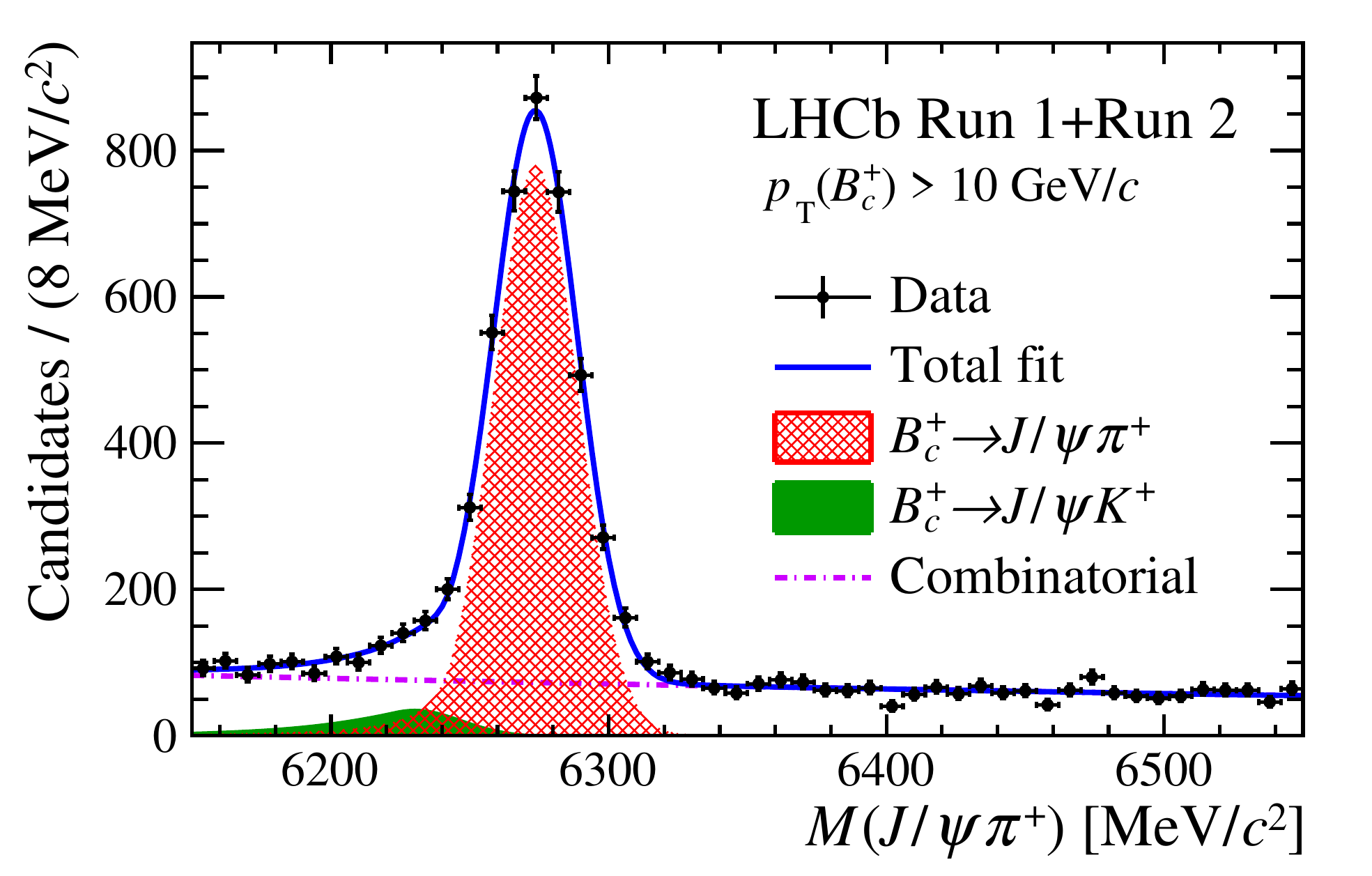}
   \caption{
   Invariant-mass distribution of the selected $\Bcp$ candidates. 
    The fit results are overlaid. 
   }
   \label{fig:Bcyield}
 \end{center}
\end{figure}

An unbinned maximum-likelihood fit is performed to 
the $M(\jpsi\pip)$ distribution.
To improve the mass resolution,
the mass $M(\jpsi\pip)$ is calculated by constraining the $\jpsi$ mass to its known value~\cite{PDG2016}
and the $\Bcp$ meson to originate from the associated PV~\cite{Hulsbergen:2005pu}.
The signal component is described by a Gaussian function with asymmetric power-law tails~\cite{Skwarnicki:1986xj}.
The parameters of the tails are determined from the simulation,
while the mean and width of the Gaussian function are left free in the fit.
The combinatorial background is modelled with an exponential function.
The contamination from the Cabibbo-suppressed decay $\Bcp \to \jpsi K^+$,
with the kaon misidentified as a pion,
is modelled by a Gaussian function with asymmetric power-law tails.
The parameters of this Gaussian function are fixed according to the simulation,
except that the mean is constrained relative to that of the $\Bcp \to \jpsi\pip$ signal.
The invariant-mass distribution
of the $\jpsi\pip$ candidates is shown in Fig.~\ref{fig:Bcyield}.
The $\Bcp$ signal yield is $3785\pm73$.
The fitted $\Bcp$ mass and mass resolution are $6273.7 \pm 0.3 \mevcc$ and
$15.1 \pm 0.3 \mevcc$, respectively.

To reconstruct the $\twoBctwos$ candidates,
$\Bcp$ candidates with \mbox{$M(\jpsi\pip)\in[6200,6320]\mevcc$} are combined with a pair of oppositely charged particles identified as pions.
These pion candidates are required to originate from the PV,
and each have $\pt>300\mevc$,
$p>1500\mevc$,
and a good track-fit quality.
The $\twoBctwos$ candidate is required to have a good vertex-fit quality.
To improve the mass resolution, 
a fit~\cite{Hulsbergen:2005pu} is performed in which the $\jpsi$ and $\Bcp$ masses are constrained to their known values~\cite{PDG2016} and the daughters of the $\twoBctwos$ meson are required to point to the associated PV.
The $\chisq$ per number of degrees of freedom of this fit must be smaller than nine.
The value of $M(\Bcp\pip\pim)-M(\Bcp)-M(\pip\pim)$
is required to be smaller than $200\mevcc$.
To ensure that the selection does not produce any 
artificial peaks in the 
$M(\Bcp\pip\pim)$ spectrum, the same requirements are applied to a same-sign
sample, constructed from $\Bcp\pip\pip$ or $\Bcp\pim\pim$ combinations.
The efficiency of the selections is found to change smoothly with 
the invariant mass $M(\Bcp\pi\pi)$ 
and no peaks are seen in the same-sign sample.

\begin{figure}[tb]
 \begin{center}
    \includegraphics*[width=.49\textwidth]{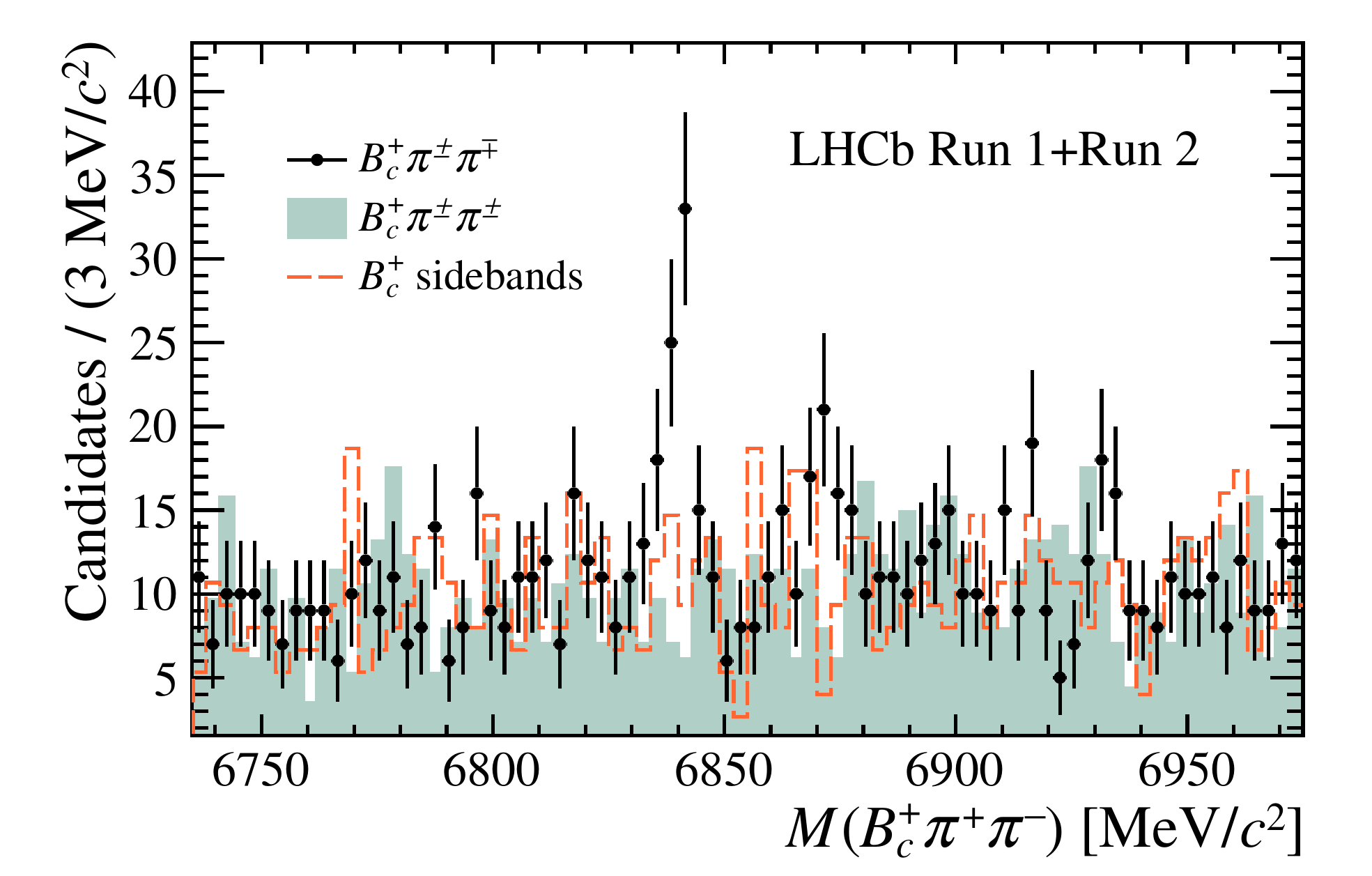}
    \caption{Invariant-mass $M(\Bcp\pip\pim)$ distributions for the data and same-sign samples
    with the distribution of the $\Bcp$ mass sidebands overlaid.
   }
   \label{fig:Bc2Smass}
 \end{center}
\end{figure}

The $M(\Bcp\pip\pim)$ distribution in the data sample
after all the selections are applied
is shown in Fig.~\ref{fig:Bc2Smass},
with those of the same-sign sample
and a sample drawn from the $\Bcp$ sidebands
($M(\jpsi\pip) \in [6150,6200] \cup [6320,6550] \mevcc$)
superimposed for comparison.
The same-sign and $\Bcp$ mass sideband distributions are 
scaled to the opposite-sign distribution in the sideband region, 
$M(\Bcp\pip\pim)\in [6735,6825] \cup [6895,6975]\mevcc$.
The $M(\Bcp\pip\pim)$ distribution presents an obvious peak at approximately $6840\mevcc$, and a less significant structure at about $6870\mevcc$.

The masses and yields of the $\twoBctwos$ peaks are determined using
an unbinned maximum-likelihood fit to the distribution of
the mass difference, 
\mbox{$\Delta M \equiv M(\Bcp\pip\pim) - M(\Bcp)$},
to eliminate the dependence on the reconstructed $\Bcp$ mass.
Here the mass $M(\Bcp\pip\pim)$ is calculated 
with no constraint on the $\Bcp$ mass,
but only constraining the $\jpsi$ mass to its known value~\cite{PDG2016}
and requiring the $\twoBctwos$ meson to come from the associated PV~\cite{Hulsbergen:2005pu}.
Each $\twoBctwos$ peak is modelled by a Gaussian function with asymmetric
power-law tails~\cite{Skwarnicki:1986xj}.
The tail parameters are fixed to the values determined from simulation,
while the Gaussian mean and width are treated as free parameters.
The combinatorial background is described by a second-order polynomial function.

\begin{figure}[tb]
 \begin{center}
    \includegraphics*[width=.49\textwidth]{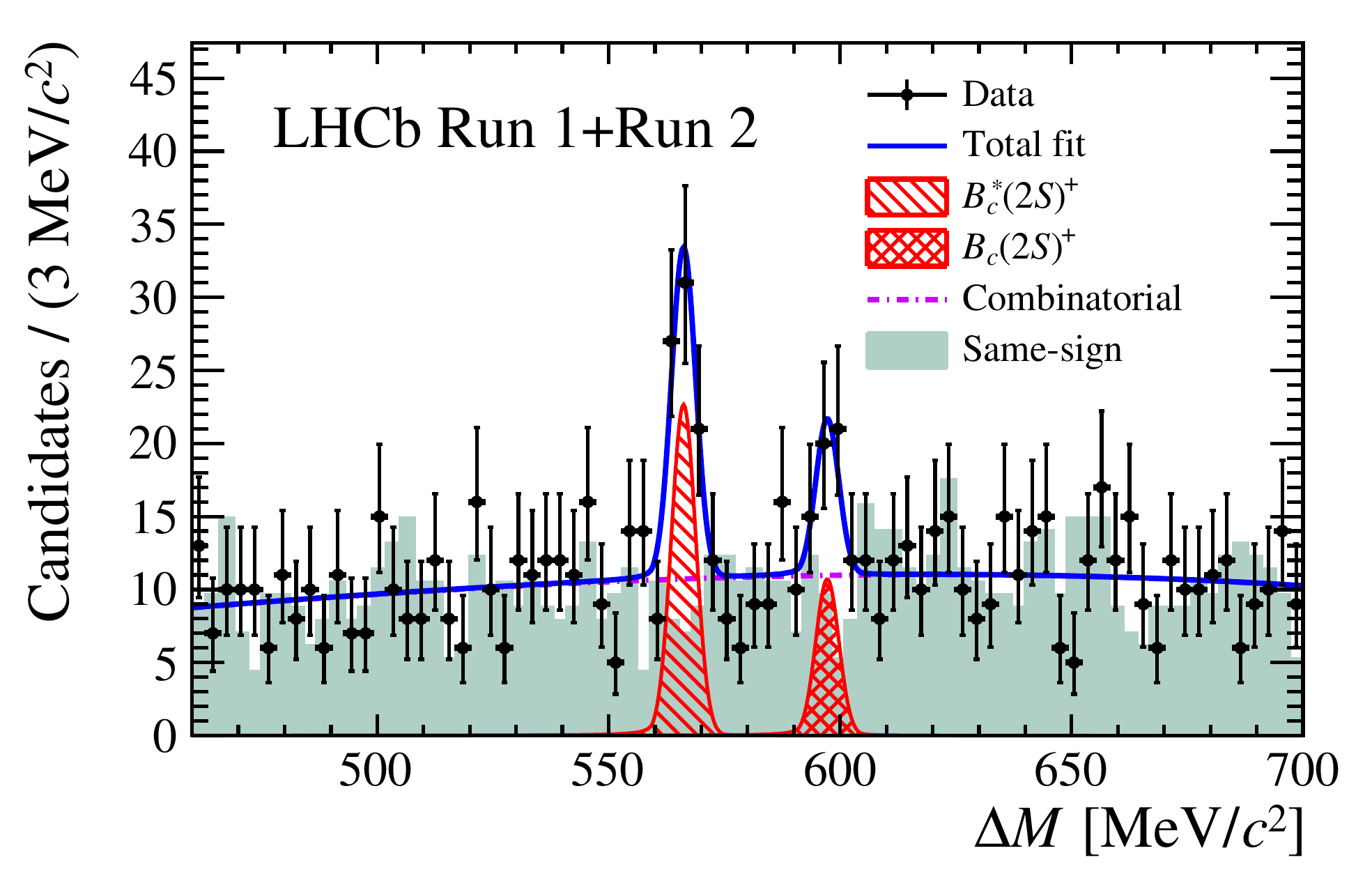}
   \caption{Distribution of $\Delta M = M(\Bcp\pip\pim) - M(\Bcp)$ with the fit results overlaid.
   The same-sign distribution has been normalized to the data in the $\twoBctwos$ sideband region.
   }
   \label{fig:fitBc2Sdata}
 \end{center}
\end{figure}

The fit to the $\Delta M$ distribution is shown in Fig.~\ref{fig:fitBc2Sdata}, and
the results are summarised in Table~\ref{tab:fit}.
The $\Bctwosstar$ signal yield is determined to be $51\pm10\stat$,
corresponding to a local statistical significance of $6.8\,\sigma$.
The significance is evaluated with a likelihood-based test,
in which the likelihood distribution of the background-only hypothesis is obtained using pseudoexperiments~\cite{Cowan:2010js}.
The yield of the $\Bctwos$ state is $24\pm9\stat$
with a local statistical significance of $3.2\,\sigma$.
The Gaussian widths of the two peaks are 
consistent with the expectation of negligible resonance widths.
The mass difference between the two peaks is measured to be $\statmgap$.
Taking the known $\Bcp$ mass,
$M(\Bcp) = 6274.9\pm0.8\mevcc$~\cite{PDG2018},
the quantities $\Mrec$ and $\MBctwos$ are determined to be 
$\statmone$ and $\statmtwo$,
respectively.
The second uncertainty is due to the limited knowledge of the
$\Bcp$ mass.
After considering the look-elsewhere effect
in the predicted mass regions~\cite{Gross:2010qma}, 
$M(\Bcp\pip\pim) \in [6790,6895]\mevcc$ for the $\Bctwosstar$
state,
and $M(\Bcp\pip\pim) \in [6845,6895]\mevcc$
for the $\Bctwos$ state~\cite{
Gershtein:1987jj, Chen:1992fq, Eichten:1994gt, Kiselev:1994rc,
Gupta:1995ps, Fulcher:1998ka, Ebert:2002pp, Godfrey:2004ya, Wei:2010zza, Rai:2006dt, El:2005},
the global statistical significances of the two states are determined to be 
$6.3\,\sigma$ and $2.2\,\sigma$, respectively.

\begin{table}[tb]
\begin{center}
\caption{\small
\label{tab:fit}
Results of the fit to the $\Delta M$ distribution.
Uncertainties are statistical only.}
\begin{tabular}{@{}lcc@{}}
\hline \hline
 & $\Bctwosstar$ & $\Bctwos$ \\
\hline
Signal yield & $~~~51\pm10$ & $~\,24\pm9$ \\
Peak $\Delta M$ value~(${\rm Me\kern -0.1em V\!/}c^2$) & $566.2\pm0.6$ & $597.2\pm1.3$ \\
Resolution~(${\rm Me\kern -0.1em V\!/}c^2$) & $~~~2.6\pm0.5$ & $~~~2.5\pm1.0$ \\
Local significance    & $6.8\,\sigma$ & $3.2\,\sigma$ \\
Global significance  & $6.3\,\sigma$ & $2.2\,\sigma$ \\
  \hline
  \hline
\end{tabular}
\end{center}
\end{table}

Several sources of systematic uncertainty on the determination of
the mass difference $\Delta M$ are studied.
The dominant contribution is from the uncertainty on the momentum scale,
which is due to 
imperfections in the description of the magnetic field
and the imperfect alignment of the subdetectors.
The uncertainty of the momentum calibration is estimated 
using other particles, such as $\KS$ and $\PUpsilon$ mesons, and
leads to an uncertainty of $0.12\mevcc$ on the $\Delta M$ measurements.  
The unreconstructed photon emitted in the $\Bctwosstar$ decay chain 
could be an additional source of systematic uncertainty. 
Studies on simulated events show that the missing photon introduces a small bias, 
and a correction of $+0.08\mevcc$, 
with negligible uncertainty, 
is applied to the fitted value of the $\Bctwosstar$ mass peak.
All other systematic uncertainties are negligible and are briefly described as follows.
The effects of the imperfect modelling of the signal and background components
are estimated by using alternative models.
The alternative model for the signal peaks uses Hypatia functions~\cite{Santos:2013gra}, 
while for the background
the alternative model consists of a sum of two threshold functions, 
each of the form $(\Delta M - m_t)^{p}\times e^{-C\cdot(\Delta M - m_t)}$, 
where $p$ and $C$ are free parameters,
and $m_t$ represents the threshold,
which is taken to be $2m_{\pi^{\pm}}$.
The changes in $\Delta M$ obtained with the alternative models are found to be negligible.
The effect of final-state radiation is also studied with simulated events 
and the associated uncertainty on the fitted mass values is found to be negligible.
The total systematic uncertainty on $\Delta M$ for both the $\Bctwos$ and $\Bctwosstar$ states
of $0.12\mevcc$
is fully correlated, and
therefore cancels in the mass difference of the two peaks.

In conclusion,
using $pp$ collision data collected by the \lhcb experiment
at centre-of-mass energies of $\sqs = 7$, $8$ and $13\tev$,
corresponding to an integrated luminosity of $8.5\invfb$,
a peaking structure consistent with the $\Bctwosstar$ state is observed in the $\Bcp\pip\pim$ mass spectrum
with a global~(local) statistical significance of $6.3\,\sigma$~($6.8\,\sigma$).
The mass associated with the $\Bctwosstar$ state, 
for which the low-energy photon in the intermediate
decay $\Bcstar \to \Bc\gamma$ is not reconstructed, 
is measured to be 
\begin{equation*}
    \mone,
\end{equation*}
where the last uncertainty is due to the limited knowledge of the $\Bcp$ mass.
It is equal to 
$\Mrec = M(\Bctwosstar) - \left( M(\Bcstar) - M(\Bcp) \right)$.
The mass difference between the $\Bctwosstar$ and $\Bcstar$ state is determined to be $\Qone$.
The data also show a hint for a second structure consistent with the $\Bctwos$ state
with a global~(local) statistical significance of $2.2\,\sigma$~($3.2\,\sigma$).
Assuming this peak is due to the $\Bctwos$ state, 
its mass is measured to be
\begin{equation*}
    \mtwo.
\end{equation*}
The mass difference between the $\Bctwos$ and $\Bcp$ state is $\Qtwo$.
The mass difference of the two $\twoBctwos$ peaks is determined to be 
\begin{equation*}
    \mgap,
\end{equation*}
in which both the uncertainty from the $\Bcp$ mass and the systematic uncertainty cancel.
The mass measurements are the most precise to date,
and are consistent with the results from the CMS collaboration~\cite{Sirunyan:2019osb}.
They are also within the range of the theoretical predictions~\cite{Gershtein:1987jj,
Chen:1992fq, Eichten:1994gt, Kiselev:1994rc,
Gupta:1995ps, Fulcher:1998ka, Ebert:2002pp, Godfrey:2004ya, Wei:2010zza, El:2005,
Davies:1996gi, Zeng:1994vj, Dowdall:2012ab}.

%% file: acknowledgements.tex
\section*{Acknowledgements}
%
%
\noindent
We thank Chao-Hsi~Chang and Xing-Gang~Wu for frequent and interesting
discussions on the production of the $B_c$ mesons.
We express our gratitude to our colleagues in the CERN
accelerator departments for the excellent performance of the LHC. We
thank the technical and administrative staff at the LHCb
institutes.
We acknowledge support from CERN and from the national agencies:
CAPES, CNPq, FAPERJ and FINEP (Brazil); 
MOST and NSFC (China); 
CNRS/IN2P3 (France); 
BMBF, DFG and MPG (Germany); 
INFN (Italy); 
NWO (Netherlands); 
MNiSW and NCN (Poland); 
MEN/IFA (Romania); 
MSHE (Russia); 
MinECo (Spain); 
SNSF and SER (Switzerland); 
NASU (Ukraine); 
STFC (United Kingdom); 
NSF (USA).
We acknowledge the computing resources that are provided by CERN, IN2P3
(France), KIT and DESY (Germany), INFN (Italy), SURF (Netherlands),
PIC (Spain), GridPP (United Kingdom), RRCKI and Yandex
LLC (Russia), CSCS (Switzerland), IFIN-HH (Romania), CBPF (Brazil),
PL-GRID (Poland) and OSC (USA).
We are indebted to the communities behind the multiple open-source
software packages on which we depend.
Individual groups or members have received support from
AvH Foundation (Germany);
EPLANET, Marie Sk\l{}odowska-Curie Actions and ERC (European Union);
ANR, Labex P2IO and OCEVU, and R\'{e}gion Auvergne-Rh\^{o}ne-Alpes (France);
Key Research Program of Frontier Sciences of CAS, CAS PIFI, and the Thousand Talents Program (China);
RFBR, RSF and Yandex LLC (Russia);
GVA, XuntaGal and GENCAT (Spain);
the Royal Society
and the Leverhulme Trust (United Kingdom);
Laboratory Directed Research and Development program of LANL (USA).

%% file: LHCb_Authorship_flat_07-Feb-2019.tex
\centerline
{\large\bf LHCb collaboration}
\begin
{flushleft}
\small
R.~Aaij$^{29}$,
C.~Abell{\'a}n~Beteta$^{46}$,
B.~Adeva$^{43}$,
M.~Adinolfi$^{50}$,
C.A.~Aidala$^{77}$,
Z.~Ajaltouni$^{7}$,
S.~Akar$^{61}$,
P.~Albicocco$^{20}$,
J.~Albrecht$^{12}$,
F.~Alessio$^{44}$
% ...

{\footnotesize \it

% ...
$ ^{7}$Universit{\'e} Clermont Auvergne, CNRS/IN2P3, LPC, Clermont-Ferrand, France\\
% ...
$ ^{12}$Fakult{\"a}t Physik, Technische Universit{\"a}t Dortmund, Dortmund, Germany\\
% ...
$ ^{20}$INFN Laboratori Nazionali di Frascati, Frascati, Italy\\
% ...
$ ^{29}$Nikhef National Institute for Subatomic Physics, Amsterdam, Netherlands\\
% ...
$ ^{43}$Instituto Galego de F{\'\i}sica de Altas Enerx{\'\i}as (IGFAE), Universidade de Santiago de Compostela, Santiago de Compostela, Spain\\
$ ^{44}$European Organization for Nuclear Research (CERN), Geneva, Switzerland\\
% ...
$ ^{46}$Physik-Institut, Universit{\"a}t Z{\"u}rich, Z{\"u}rich, Switzerland\\
% ...
$ ^{50}$H.H. Wills Physics Laboratory, University of Bristol, Bristol, United Kingdom\\
% ...
$ ^{61}$University of Cincinnati, Cincinnati, OH, United States\\
% ...
$ ^{63}$Syracuse University, Syracuse, NY, United States\\
% ...
$ ^{77}$University of Michigan, Ann Arbor, United States, associated to $^{63}$\\
% ...
}
\end{flushleft}